\documentclass[11pt]{article}
\usepackage[utf8]{inputenc}
\usepackage{amsfonts}
\usepackage{amsmath}
\usepackage{amssymb}
\usepackage{indentfirst}
\usepackage{graphicx}
\usepackage{hyperref}
\usepackage{cite}

\numberwithin{equation}{section}
\begin{document}

\begin{titlepage}
\vspace{3cm}

\baselineskip=24pt

\begin{center}
\textbf{\LARGE{Generalized Maxwellian exotic Bargmann gravity theory in three spacetime dimensions}}
\par\end{center}{\LARGE \par}

\begin{center}
	\vspace{1cm}
	\textbf{Patrick Concha}$^{\ast}$,
    \textbf{Marcelo Ipinza}$^{\ddag}$,
	\textbf{Evelyn Rodríguez}$^{\dag}$,
	\small
	\\[5mm]
    $^{\ast}$\textit{Departamento de Matemática y Física Aplicadas, }\\
	\textit{ Universidad Católica de la Santísima Concepción, }\\
\textit{ Alonso de Ribera 2850, Concepción, Chile.}
	\\[2mm]
	$^{\ddag}$\textit{Instituto
		de Física, Pontificia Universidad Católica de Valparaíso, }\\
	\textit{ Casilla 4059, Valparaiso-Chile.}
	\\[2mm]
	$^{\dag}$\textit{Departamento de Ciencias, Facultad de Artes Liberales,} \\
	\textit{Universidad Adolfo Ibáñez, Viña del Mar-Chile.} \\[5mm]
	\footnotesize
	\texttt{patrick.concha@ucsc.cl},
    \texttt{marcelo.calderon@pucv.cl},
	\texttt{evelyn.rodriguez@edu.uai.cl},
	\par\end{center}
\vskip 26pt
\begin{abstract}
\noindent We present a generalization of the so-called Maxwellian extended Bargmann algebra by considering a non-relativistic limit to a generalized Maxwell algebra defined in three spacetime dimensions. The non-relativistic Chern-Simons gravity theory based on this new algebra is also constructed and discussed. We point out that the extended Bargmann and its Maxwellian generalization are particular sub-cases of the generalized Maxwellian extended Bargmann gravity introduced here. The extension of our results using the semigroup expansion method is also discussed.

\end{abstract}
\end{titlepage}\newpage {} {\baselineskip=12pt \tableofcontents{}}

\section{Introduction}

There has been a growing interest in exploring non-relativistic (NR) gravity
theories \cite{Kuchar, DK, DBKP, DGH, Duval, DLP, DLP2, Horava, DH, ABPR,
ABGR, ABRS, BMM, BM1, BM2, BRZ2, BMu, BCRR, HHO, BMu2, OOTZ, MPS, CS, CRR3,
CRR2, GKPSR, BGSR, EHVdB}. In three spacetime dimensions, gravity models can
be formulated using the Chern-Simons (CS) formalism \cite{AT, Witten,
Zanelli} offering a simpler framework to construct non-relativistic gravity
actions. Furthermore, three-dimensional CS gravity theories can be seen as
interesting toy models to approach higher-dimensional theories.

The construction of a proper finite NR CS action without degeneracy may
require to enlarge the field content of the relativistic theory \cite{BRZ,
GO, BCG}. In the case of three-dimensional Einstein gravity without
cosmological constant, it is necessary to consider two additional $U\left(
1\right) $ gauge fields in order to define a consistent NR\ limit leading to
the so-called extended Bargmann gravity \cite{BR, HLO}. The incorporation of
a cosmological constant modifies the theory to the so-called extended
Newton-Hooke gravity \cite{PS, AMO, Gao, GP, BGK, AGKP, OOZ}. More recently,
a NR version of a three-dimensional gravity theory coupled to
electromagnetism has been presented in \cite{AFGHZ} describing what they
called as Maxwellian extended Bargmann (MEB) gravity. Such NR theory
requires to introduce three extra $U\left( 1\right) $ gauge fields to the
Maxwell algebra.

The Maxwell algebra has been introduced in \cite{Schrader, BCR, GK}\ in
order to describe a Minkowski space in the presence of a electromagnetic
field background. In three spacetime dimensions, a CS gravity action without
cosmological constant invariant under the Maxwell algebra has been presented
in \cite{SSV, HR, CFRS} whose general solution and asymptotic structure have
been studied in \cite{CMMRSV}. More recently, an isomorphic (dual) version
of the Maxwell algebra denoted as Hietarinta-Maxwell algebra has been of
particular interest for exploring spontaneous breaking of symmetry \cite{BS,
CSDS}.

A generalization of the Maxwell algebra has been introduced in \cite{EHTZ,
GRCS}\ and has been denoted as $\mathfrak{B}_{5}$ algebra. This
generalization is characterized by the presence of an additional generator
with respect to the Maxwell algebra. Interestingly, the aforesaid algebra
belongs to a larger family of algebras denoted as $\mathfrak{B}_{k}$ where $%
\mathfrak{B}_{4}$ and $\mathfrak{B}_{3}$ are the Maxwell and Poincaré
algebras, respectively. Such family has been useful to recover standard
General Relativity without cosmological constant from a CS\ and Born-Infeld
gravity theory \cite{GRCS, CPRS1, CPRS2, CPRS3}. Subsequently, the coupling
of spin-3 gauge field to $\mathfrak{B}_{k}$ CS gravity models in three
spacetime dimensions has been explored in \cite{CCFRS}.

It is natural to address the question whether such generalized Maxwell
algebra admits a well-defined NR version in three spacetime dimensions. Here
we show that the relativistic theory has to be enlarged with four $U\left(
1\right) $ gauge fields in order to apply an Inönü-Wigner (IW) contraction
\cite{IW, WW} leading to a non-degenerate and finite NR CS gravity theory.
The new symmetry obtained corresponds to a generalization of the MEB algebra
and has been called GMEB\ algebra.

An alternative way to find the GMEB symmetry is also discused considering
the semigroup expansion method ($S$-expansion) \cite{Sexp, CKMN, AMNT,
ACIPS, ILPR} and following the procedure used in \cite{PSR}. As was shown in
\cite{PSR}, a generalized family of NR algebras, that we have denoted as
generalized extended Bargmann algebra, can be obtained using the $S$%
-expansion procedure. In particular, we show that the extended Bargmann, the
MEB and the GMEB algebras are particular sub-cases of this family of NR
algebras. The expansion procedure considered here can be seen as a general
method allowing to classify diverse NR symmetries by providing the proper
NR\ limit and the additional gauge fields required in the relativistic
theory. Interestingly, the $S$-expansion method provides not only with the
commutation relations of the new NR algebras but also with the non-vanishing
components of the invariant tensors which are essential to the construction\
of NR CS actions.

The paper is organized as follows: in Section 2, we give a brief review of
the generalized Maxwell algebra. The corresponding relativistic CS action
and its $U\left( 1\right) $ enlargement are also presented. Sections 3 and 4
contain our main results. In particular, in Section 3 we present the
contraction process leading to the GMEB gravity theory. The family of NR
algebras obtained through the semigroup expansion procedure is presented in
section 4. Section 5 concludes our work with some discussion about possible
future developments.

\section{Relativistic gravity and generalized Maxwell algebra}

In this section we briefly review the generalized Maxwell algebra and
present the construction of a three-dimensional CS gravity action invariant
under such algebra. \

A generalization of the Maxwell algebra has been first introduced as the $%
\mathfrak{B}_{5}$ algebra in \cite{EHTZ, GRCS}. It is characterized by the
presence of the spacetime rotations $J_{A}$, the spacetime translations $%
P_{A}$, the so-called Maxwell gravitational generator $Z_{A}$ and a new type
of generator that we have denoted as $N_{A}$. The generators of the
generalized Maxwell algebra satisfy the following non-vanishing commutation
relations:%
\begin{eqnarray}
\left[ J_{A},J_{B}\right] &=&\epsilon _{ABC}J^{C}\,,\text{ \ \ \ \ }\left[
P_{A},P_{B}\right] =\epsilon _{ABC}Z^{C}\,,  \notag \\
\left[ J_{A},P_{B}\right] &=&\epsilon _{ABC}P^{C}\,,\text{ \ \ \ \ }\left[
J_{A},N_{B}\right] =\epsilon _{ABC}N^{C}\,,  \label{gMAX} \\
\left[ J_{A},Z_{B}\right] &=&\epsilon _{ABC}Z^{C}\,,\text{ \ \ \ \ }\left[
Z_{A},P_{B}\right] =\epsilon _{ABC}N^{C}\,,  \notag
\end{eqnarray}%
where $A,B,C=0,1,2$ are the Lorentz indices which are raised and lowered
with the Minkowski metric. Here $\epsilon _{ABC}$ corresponds to the Levi
Civita tensor which satisfies $\epsilon _{012}=-\epsilon ^{012}=1$. It is
interesting to point out that the commutator $\left[ P_{a},P_{b}\right] $ is
proportional to the Maxwell gravitational generator $Z_{A}$ as in the
Maxwell algebra. Nevertheless, the commutator $\left[ Z_{A},P_{B}\right] $
is no longer zero due to the presence of the new generator $N_{A}$.
Furthermore, unlike the AdS-Lorentz algebra \cite{SS, DFIMRSV, SaSa}, this
generalization is not a deformation of the Maxwell algebra and then does not
reproduce the Maxwell symmetry through a contraction process.

Although this algebra and its generalizations have been explored with
diverse applications, a three-dimensional CS gravity action based on this
generalization of the Maxwell algebra has not been explicitly presented. A
CS action in three spacetime dimensions reads%
\begin{equation}
I\left[ A\right] =\int \langle AdA+\frac{2}{3}\,A^{3}\rangle \,,
\label{CSaction}
\end{equation}%
where $\left\langle \cdots \right\rangle $ denotes the invariant trace and $%
A=A^{a}T_{a}$ corresponds to the gauge connection one-form. In our case, the
connection one-form $A$ is given by%
\begin{equation}
A=W^{A}J_{A}+E^{A}P_{A}+K^{A}Z_{A}+U^{A}N_{A}\,,  \label{Conn}
\end{equation}%
where $W^{A}$ is the spin connection one-form, $E^{A}$ is the vielbein, $%
K^{A}$ is the so-called gravitational Maxwell gauge field and $U^{A}$ is the
new gauge field along the Abelian generator $N_{A}$. The respective
curvature two form $F=dA+\frac{1}{2}\left[ A,A\right] $ reads%
\begin{equation}
F=R^{A}(W)J_{A}+R^{A}(E)P_{A}+R^{A}(K)Z_{A}+R^{A}\left( U\right) N_{A}\,,
\end{equation}%
where the Lorentz curvature $R^{A}\left( W\right) $, the torsion $%
R^{A}\left( E\right) $ and the curvatures along the generators $Z_{A}$ and $%
N_{A}$ are respectively given by%
\begin{eqnarray}
R^{A}(W):= &&dW^{A}-\frac{1}{2}\epsilon ^{ABC}W_{B}W_{C}\,,  \notag \\
R^{A}(E):= &&D_{W}E^{A}\,,  \notag \\
R^{A}(K):= &&D_{W}K^{A}\,-\frac{1}{2}\epsilon ^{ABC}E_{B}E_{C}\,, \\
R^{A}\left( U\right) := &&D_{W}U^{A}-\epsilon ^{ABC}K_{B}E_{C}\,.  \notag
\end{eqnarray}%
Here $D_{W}\Theta ^{A}:=d\Theta ^{A}-\epsilon ^{ABC}W_{B}\Theta _{C}$ is the
usual Lorentz covariant derivative.

In order to construct the relativistic CS gravity action invariant under the
algebra (\ref{gMAX}) we shall consider the most general non-vanishing
components of the invariant tensor \cite{GRCS}%
\begin{eqnarray}
\left\langle J_{A}J_{B}\right\rangle &=&\alpha _{0}\,\eta _{AB}\,,\text{ \ \
\ }\left\langle P_{A}P_{B}\right\rangle =\alpha _{2}\,\eta _{AB}\,,  \notag
\\
\left\langle J_{A}P_{B}\right\rangle &=&\alpha _{1}\,\eta _{AB}\,,\text{ \ \
\thinspace }\left\langle J_{A}N_{B}\right\rangle =\alpha _{3}\,\eta _{AB}\,,
\label{Invten} \\
\left\langle J_{A}Z_{B}\right\rangle &=&\alpha _{2}\,\eta _{AB}\,,\text{ \ \
\thinspace }\left\langle Z_{A}P_{B}\right\rangle =\alpha _{3}\,\eta _{AB}\,,
\notag
\end{eqnarray}%
where $\alpha _{0}$, $\alpha _{1}$, $\alpha _{2}$ and $\alpha _{3}$ are
arbitrary constants. Then, considering the gauge connection one-form (\ref%
{Conn}) and the invariant tensor (\ref{Invten}) in the general form of a CS
action (\ref{CSaction}), one gets%
\begin{eqnarray}
I_{R} &=&\int \left[ \alpha _{0}\left( W^{A}dW_{A}+\frac{1}{3}\epsilon
_{ABC}W^{A}W^{B}W^{C}\right) +2\alpha _{1}E^{A}R_{A}(W)\right.  \notag \\
&&\left. +\alpha _{2}\left( 2K^{A}R_{A}\left( W\right) +E^{A}R_{A}\left(
E\right) \right) \right.  \notag \\
&&\left. +\alpha _{3}\left( 2U^{A}R_{A}\left( W\right) +2E^{A}D_{W}K^{A}+%
\frac{1}{3}\epsilon ^{ABC}E_{A}E_{B}E_{C}\right) \right] \,.  \label{RelCS}
\end{eqnarray}%
One can see that such relativistic CS action contains four independent
sectors proportional to $\alpha _{0}$, $\alpha _{1}$, $\alpha _{2}$ and $%
\alpha _{4}$. In particular, the term proportional to $\alpha _{0}$
corresponds to the so-called exotic Einstein action \cite{Witten}. The term
along $\alpha _{1}$ is the usual Einstein-Hilbert term and corresponds to
the CS action based on the Poincaré symmetry. On the other hand, the
Maxwellian gravitational gauge field has contribution in the Maxwell CS
action proportional to $\alpha _{2}$ \cite{AFGHZ, SSV, HR, CFRS, CMMRSV,
CPR, Concha} and to the new term $\alpha _{3}$. The new gauge field $U^{A}$
appears only along the $\alpha _{3}$ term together to the cosmological
constant term. Let us note that each sector is invariant under the
generalized Maxwell algebra (\ref{gMAX}). Indeed, one can show that the CS
action (\ref{RelCS}) is invariant under the following infinitesimal gauge
transformations:%
\begin{eqnarray}
\delta _{\Lambda }W^{A} &=&D_{W}\rho ^{A}\,,  \notag \\
\delta _{\Lambda }E^{A} &=&D_{W}\varepsilon ^{A}-\epsilon ^{ABC}E_{B}\rho
_{C}\,,  \notag \\
\delta _{\Lambda }K^{A} &=&D_{W}\gamma ^{A}-\epsilon ^{ABC}\left(
E_{B}\varepsilon _{C}-\rho _{B}K_{C}\right) \,, \\
\delta _{\Lambda }U^{A} &=&D_{W}\nu ^{A}-\epsilon ^{ABC}\left( E_{B}\gamma
_{C}-\rho _{B}U_{C}\right) \,,  \notag
\end{eqnarray}%
where $\Lambda =\rho ^{A}J_{A}+\varepsilon ^{A}P_{A}+\gamma ^{A}Z_{A}+\nu
^{A}N_{A}$ is the gauge parameter.

The field equations coming from (\ref{RelCS}) read%
\begin{eqnarray}
\delta W^{A}:\text{ \ \ }0 &=&\alpha _{0}R_{A}(W)+\alpha _{1}R_{A}\left(
E\right) +\alpha _{2}\left( D_{W}K_{A}-\frac{1}{2}\,\epsilon
_{ABC}E^{B}E^{C}\right)  \notag \\
&&+\alpha _{3}\left( D_{W}U_{A}-\epsilon _{ABC}K^{B}E^{C}\right) \,\,,
\notag \\
\delta E^{A}:\text{ \ \ }0 &=&\alpha _{1}R_{A}(W)+\alpha _{2}R_{A}\left(
E\right) \,+\alpha _{3}\left( D_{W}K_{A}-\frac{1}{2}\,\epsilon
_{ABC}E^{B}E^{C}\right) \,, \\
\delta K^{A}:\text{ \ \ }0 &=&\alpha _{2}R_{A}(W)+\alpha _{3}R_{A}\left(
E\right) \,\,,  \notag \\
\delta U^{A}:\text{ \ \ }0 &=&\alpha _{3}R_{A}(W)\,.  \notag
\end{eqnarray}%
which imply the vanishing of every curvature when $\alpha _{3}\neq 0$.

The study of a NR limit, as in the Maxwell and Poincaré cases, requires to
introduce $U\left( 1\right) $ gauge fields in order to avoid infinities and
cancel divergences. Such enlargement will allow to define a proper NR limit
whose NR algebra will admit non-degenerate bilinear form.

\subsection{U(1) enlargements}

Let us now consider a particular $U\left( 1\right) $ enlargement of the
generalized Maxwell algebra by adding four extra $U\left( 1\right) $
one-form gauge fields to the field content as%
\begin{equation}
A=W^{A}J_{A}+E^{A}P_{A}+K^{A}Z_{A}+U^{A}N_{A}+MY_{1}+SY_{2}+TY_{3}+VY_{4}\,.
\label{U1f}
\end{equation}%
The new relativistic algebra, [generalized Maxwell]$\oplus u\left( 1\right)
^{4}$ algebra, admits the non-vanishing components of the invariant tensor (%
\ref{Invten}) along with%
\begin{eqnarray}
\left\langle Y_{2}Y_{2}\right\rangle &=&\alpha _{0}\,,  \notag \\
\left\langle Y_{1}Y_{2}\right\rangle &=&\alpha _{1}\,,  \notag \\
\left\langle Y_{2}Y_{3}\right\rangle &=&\left\langle Y_{1}Y_{1}\right\rangle
=\alpha _{2}\,,  \label{Invten2} \\
\left\langle Y_{2}Y_{4}\right\rangle &=&\left\langle Y_{1}Y_{3}\right\rangle
=\alpha _{3}\,.  \notag
\end{eqnarray}

Considering the gauge connection one-form (\ref{U1f}) and the invariant
tensor given by (\ref{Invten}) and (\ref{Invten2}) in the general expression
of the CS action (\ref{CSaction}), we find the following relativistic CS
gravity action%
\begin{eqnarray}
I_{R} &=&\int \left[ \alpha _{0}\left( W^{A}dW_{A}+\frac{1}{3}\epsilon
_{ABC}W^{A}W^{B}W^{C}+SdS\right) +\alpha _{1}\left(
2E^{A}R_{A}(W)+2MdS\right) \right.  \notag \\
&&\left. +\alpha _{2}\left( 2K^{A}R_{A}\left( W\right) +E^{A}R_{A}\left(
E\right) +MdM+2SdT\right) \right.  \label{U1CS} \\
&&\left. +\alpha _{3}\left( 2U^{A}R_{A}\left( W\right) +2E^{A}D_{W}K^{A}+%
\frac{1}{3}\epsilon ^{ABC}E_{A}E_{B}E_{C}+2SdV+2MdT\right) \right] \,.
\notag
\end{eqnarray}

In the next section, we shall see that the presence of these abelian gauge
fields are essential to obtain, after a contraction procedure, a
well-defined NR version of the generalized Maxwell algebra without
degeneracy. Furthermore, we will show that there is a relation between the
number of $U\left( 1\right) $ generators required in the relativistic theory
and the number of elements of the semigroup involved in the semigroup
expansion method.

\section{Non-relativistic generalized Maxwell Chern-Simons gravity}

In this section, we shall consider the IW contraction of the previously
introduced relativistic algebra, and we will obtain a NR version of the
[generalized Maxwell]$\oplus u\left( 1\right) ^{4}$ algebra. Then, we will
consider the construction of a NR CS action based on the aforesaid NR
algebra. For this purpose, we will provide with the non-vanishing components
of the invariant tensor, which are derived as an IW contraction from the
relativistic invariant tensor (\ref{Invten}).

\subsection{Generalized Maxwellian exotic Bargmann algebra}

In the previous section, we have presented an $U\left( 1\right) $
enlargement of the relativistic generalized Maxwell algebra. Here, we will
obtain the NR version of this algebra. To this aim, we will introduce a
dimensionless parameter $\xi $, and we will express the relativistic
generators $\left\{
J_{0},J_{a},P_{0},P_{a},Z_{0},Z_{a},N_{0},N_{a},Y_{1},Y_{2},Y_{3},Y_{4}%
\right\} $ as a linear combination of new generators involving the $\xi $
parameter.

As in refs. \cite{AFGHZ, CR4}, we define the IW contraction process through
the identification of the relativistic generators defining the [generalized
Maxwell]$\oplus \,u(1)^{4}$ algebra, with the NR generators (denoted with a
tilde) as%
\begin{eqnarray}
J_{0} &=&\frac{\tilde{J}}{2}+\xi ^{2}\tilde{S}\,,\text{\ \ \ \ \ }J_{a}=\xi
\tilde{G}_{a}\,,\text{ \ \ \ \ \ \ }Y_{2}=\frac{\tilde{J}}{2}-\xi ^{2}\tilde{%
S}\,,  \notag \\
P_{0} &=&\frac{\tilde{H}}{2\xi }+\xi \tilde{M}\,,\text{ \ \ \ }P_{a}=\tilde{P%
}_{a}\,,\text{ \ \ \ \ \ \ \ \thinspace }Y_{1}=\frac{\tilde{H}}{2\xi }-\xi
\tilde{M}\,,  \label{con} \\
Z_{0} &=&\frac{\tilde{Z}}{2\xi ^{2}}+\tilde{T}\,,\text{ \ \ \ \ }Z_{a}=\frac{%
\tilde{Z}_{a}}{\xi }\,,\text{ \ \ \ \ \ \ \thinspace }Y_{3}=\frac{\tilde{Z}}{%
2\xi ^{2}}-\tilde{T}\,,  \notag \\
N_{0} &=&\frac{\tilde{N}}{2\xi ^{3}}+\frac{\tilde{V}}{\xi }\,,\text{ \ \ \ }%
N_{a}=\frac{\tilde{N}_{a}}{\xi ^{2}}\,,\text{ \ \ \ \ \thinspace \thinspace
\thinspace }Y_{4}=\frac{\tilde{N}}{2\xi ^{3}}-\frac{\tilde{V}}{\xi }.\,
\notag
\end{eqnarray}%
Considering this redefinition and applying the limit $\xi \rightarrow \infty
$, the contraction of the [generalized Maxwell]$\oplus u\left( 1\right) ^{4}$
algebra leads to a new NR algebra. In particular, the NR generators satisfy
the following commutation relations,%
\begin{eqnarray}
\left[ \tilde{J},\tilde{G}_{a}\right] &=&\epsilon _{ab}\tilde{G}_{b}\,,\text{
\ \ \ \ \ \ \ \ }\left[ \tilde{G}_{a},\tilde{G}_{b}\right] =-\epsilon _{ab}%
\tilde{S}\,,\text{\ \ \ \ \ \ }\,\left[ \tilde{H},\tilde{G}_{a}\right]
=\epsilon _{ab}\tilde{P}_{b}\,,\text{ \ \ }  \notag \\
\left[ \tilde{J},\tilde{P}_{a}\right] &=&\epsilon _{ab}\tilde{P}%
_{b}\,,\qquad \quad \,\,\left[ \tilde{G}_{a},\tilde{P}_{b}\right] =-\epsilon
_{ab}\tilde{M}\,\,,\text{ \ \ \ \ \ }\left[ \tilde{H},\tilde{P}_{a}\right]
=\epsilon _{ab}\tilde{Z}_{b}\,,  \notag \\
,\text{ \ \ \ \ \ \ }\left[ \tilde{J},\tilde{Z}_{a}\right] &=&\epsilon _{ab}%
\tilde{Z}_{b}\,,\qquad \quad \,\,\left[ \tilde{P}_{a},\tilde{P}_{b}\right]
=-\epsilon _{ab}\tilde{T}\,,\qquad \,\,\,\left[ \tilde{H},\tilde{Z}_{a}%
\right] =\epsilon _{ab}\tilde{N}_{b}\,,  \label{NRgen} \\
\left[ \tilde{Z},\tilde{G}_{a}\right] &=&\epsilon _{ab}\tilde{Z}%
_{b}\,,\qquad \quad \left[ \tilde{G}_{a},\tilde{Z}_{b}\right] =-\epsilon
_{ab}\tilde{T}\,,\qquad \quad \left[ \tilde{Z},\tilde{P}_{a}\right]
=\epsilon _{ab}\tilde{N}_{b}\,,  \notag \\
\text{ \ \ \ \ \ }\left[ \tilde{J},\tilde{N}_{a}\right] &=&\epsilon _{ab}%
\tilde{N}_{b}\,,\qquad \quad \left[ \tilde{P}_{a},\tilde{Z}_{b}\right]
=-\epsilon _{ab}\tilde{V}\,,\qquad \,\,\,\left[ \tilde{N},\tilde{G}_{a}%
\right] =\epsilon _{ab}\tilde{N}_{b}\,,\text{\ \ }  \notag \\
\left[ \tilde{G}_{a},\tilde{N}_{b}\right] &=&-\epsilon _{ab}\tilde{V}\,,
\notag
\end{eqnarray}%
where we have defined $\epsilon _{ab}\equiv \epsilon _{0ab},$ $\epsilon
^{ab}\equiv \epsilon ^{0ab},$ while $a=1,2$. This is a novel NR algebra
which we shall call as generalized Maxwellian extended Bargmann (GMEB)
algebra. From (\ref{NRgen}) we can see that it contains four central
extensions given by $\tilde{M}$, $\tilde{S}$, $\tilde{T}$ and $\tilde{V}$,
which are related to the four extra $U\left( 1\right) $ generators. Let us
note that the extended Bargmann algebra \cite{BR, HLO} can be recoverred by
setting $\tilde{Z}=\tilde{Z}_{a}=\tilde{T}=\tilde{N}=\tilde{N}_{a}=\tilde{V}%
=0$. On the other hand, if we set $\tilde{N}=\tilde{N}_{a}=\tilde{V}=0$, the
Maxwellian Extended Bargmann algebra is obtained \cite{AFGHZ}. As we shall
see, the presence of the central charges assures to have non-degenerate
invariant bilinear form.

\subsection{Non-relativistic generalized Maxwell Chern-Simons action}

In order to construct a CS action for the GMEB algebra we need the NR
invariant tensor. The diverse components can be obtained from the
contraction (\ref{con}) of the relativistic invariant tensor (\ref{Invten}).
The non-vanishing components of a non-degenerate invariant tensor for the
GMEB algebra are given by%
\begin{eqnarray}
\left\langle \tilde{J}\tilde{S}\right\rangle &=&-\tilde{\alpha}_{0}\,,\text{
\ \ \ \ \ \ \ \ \ \ \ \ \ \ \ \ \ \ \ \ \ \ \ \ \ \ }  \notag \\
\left\langle \tilde{G}_{a}\tilde{G}_{b}\right\rangle &=&\tilde{\alpha}%
_{0}\delta _{ab}\,,  \notag \\
\left\langle \tilde{G}_{a}\tilde{P}_{b}\right\rangle &=&\tilde{\alpha}%
_{1}\delta _{ab}\,,  \notag \\
\left\langle \tilde{H}\tilde{S}\right\rangle &=&\left\langle \tilde{M}\tilde{%
J}\right\rangle =-\tilde{\alpha}_{1}\,,  \notag \\
\left\langle \tilde{P}_{a}\tilde{P}_{b}\right\rangle &=&\left\langle \tilde{G%
}_{a}\tilde{Z}_{b}\right\rangle =\tilde{\alpha}_{2}\delta _{ab}\,,
\label{nrinv} \\
\left\langle \tilde{J}\tilde{T}\right\rangle &=&\left\langle \tilde{H}\tilde{%
M}\right\rangle =\left\langle \tilde{S}\tilde{Z}\right\rangle =-\tilde{\alpha%
}_{2}\,,  \notag \\
\left\langle \tilde{G}_{a}\tilde{N}_{b}\right\rangle &=&\left\langle \tilde{P%
}_{a}\tilde{Z}_{b}\right\rangle =\tilde{\alpha}_{3}\delta _{ab}\,,  \notag \\
\left\langle \tilde{J}\tilde{V}\right\rangle &=&\left\langle \tilde{H}\tilde{%
T}\right\rangle =\left\langle \tilde{M}\tilde{Z}\right\rangle =\left\langle
\tilde{S}\tilde{N}\right\rangle =-\tilde{\alpha}_{3}\,,  \notag
\end{eqnarray}%
where the relativistic parameters $\alpha $'s have been rescaled as
\begin{equation}
\alpha _{0}=\tilde{\alpha}_{0}\xi ^{2}\,,\text{ \ \ \ \ }\alpha _{1}=\tilde{%
\alpha}_{1}\xi \,,\text{ \ \ \ \ }\alpha _{2}=\tilde{\alpha}_{2}\,,\text{ \
\ \ \ }\alpha _{3}=\tilde{\alpha}_{3}\xi ^{-1}.  \label{alphas}
\end{equation}%
As in \cite{AFGHZ, CR4}, such rescaling is done in order to have a finite NR
CS action. \ Now we are ready to construct the aforesaid CS action. The NR
one-form gauge connection $\tilde{A}$ reads%
\begin{eqnarray}
\tilde{A} &=&\tau \tilde{H}+e^{a}\tilde{P}_{a}+\omega \tilde{J}+\omega ^{a}%
\tilde{G}_{a}+k\tilde{Z}+k^{a}\tilde{Z}_{a}+f\tilde{N}  \notag \\
&&+f^{a}\tilde{N}_{a}+m\tilde{M}+s\tilde{S}+t\tilde{T}\,+v\tilde{V}.
\label{oneform}
\end{eqnarray}%
The corresponding NR curvature two-form is then written as%
\begin{eqnarray}
\tilde{F} &=&R(\tau )\tilde{H}+R^{a}(e^{b})\tilde{P}_{a}+R(\omega )\tilde{J}%
+R^{a}(\omega ^{b})\tilde{G}_{a}+R(k)\tilde{Z}+R^{a}(k^{b})\tilde{Z}_{a}
\notag \\
&&+R\left( f\right) \tilde{N}+R^{a}\left( f^{b}\right) \tilde{N}_{a}+R(m)%
\tilde{M}+R(s)\tilde{S}+R(t)\tilde{T}+R\left( v\right) \tilde{V}\,,
\end{eqnarray}%
where%
\begin{eqnarray}
R(\tau ) &=&d\tau \,,  \notag \\
R^{a}(e^{b}) &=&de^{a}+\epsilon ^{ac}\omega e_{c}+\epsilon ^{ac}\tau \omega
_{c}\,,  \notag \\
R(\omega ) &=&d\omega \,,  \notag \\
R^{a}(\omega ^{b}) &=&d\omega ^{a}+\epsilon ^{ac}\omega \omega _{c}\,,
\notag \\
R(k) &=&dk\,,  \notag \\
R^{a}(k^{b}) &=&dk^{a}+\epsilon ^{ac}\omega k_{c}+\epsilon ^{ac}\tau
e_{c}+\epsilon ^{ac}k\omega _{c}\,,  \label{Curva} \\
R^{a}\left( f^{b}\right) &=&df^{a}+\epsilon ^{ac}\omega f_{c}+\epsilon
^{ac}\tau k_{c}+\epsilon ^{ac}ke_{c}+\epsilon ^{ac}f\omega _{c}\,,  \notag \\
R\left( f\right) &=&df\,,  \notag \\
R(m) &=&dm+\epsilon ^{ac}e_{a}\omega _{c}\,,  \notag \\
R(s) &=&ds+\frac{1}{2}\epsilon ^{ac}\omega _{a}\omega _{c}\,,  \notag \\
R(t) &=&dt+\epsilon ^{ac}\omega _{a}k_{c}+\frac{1}{2}\epsilon
^{ac}e_{a}e_{c}\,,  \notag \\
R\left( v\right) &=&dv+\epsilon ^{ac}\omega _{a}f_{c}+\epsilon
^{ac}e_{a}k_{c}\,.  \notag
\end{eqnarray}%
The NR CS action invariant under the GMEB algebra can be computed by
replacing\ the NR one-form connection (\ref{oneform}) and the invariant
tensor (\ref{nrinv}) in the general expression for the CS action in three
spacetime (\ref{CSaction}), or by taking the NR limit directly in (\ref{U1CS}%
). In both cases, the resulting NR CS action is given by%
\begin{eqnarray}
I_{NR} &=&\int \tilde{\alpha}_{0}\left[ \omega _{a}R^{a}(\omega
^{b})-2sR\left( \omega \right) \right] +\tilde{\alpha}_{1}\left[
2e_{a}R^{a}(\omega ^{b})-2mR(\omega )-2\tau R(s)\right]  \notag \\
&&+\tilde{\alpha}_{2}\left[ e_{a}R^{a}\left( e^{b}\right) +k_{a}R^{a}\left(
\omega ^{b}\right) +\omega _{a}R^{a}\left( k^{b}\right) -2sR\left( k\right)
-2mR\left( \tau \right) -2tR\left( \omega \right) \right]  \label{NRaction}
\\
&&\tilde{\alpha}_{3}\,\left[ \omega _{a}R^{a}\left( f^{b}\right)
+f_{a}R^{a}\left( \omega ^{b}\right) +e_{a}R^{a}\left( k^{b}\right)
+k_{a}R^{a}\left( e^{b}\right) -2sR\left( f\right) \right.  \notag \\
&&\left. -2vR\left( \omega \right) -2mR\left( k\right) -2tR\left( \tau
\right) \right] \,.  \notag
\end{eqnarray}%
From (\ref{NRaction}), we see that it contains four independent sectors,
each one of those proportional to an arbitrary constant $\tilde{\alpha}_{i}.$
The first term corresponds to the NR version of the Exotic gravity \cite%
{Witten} which is denoted as NR exotic gravity. The second term proportional
to $\tilde{\alpha}_{1}$ reproduces the extended Bargmann gravity action \cite%
{BR, HLO}, while the third term is the MEB gravity action introduced in \cite%
{AFGHZ}. The new gauge fields $f_{a}$, $f$ and $v$, appear explicitly in the
last term proportional to $\tilde{\alpha}_{3}$, which corresponds to the CS
action for the new NR generalized Maxwell algebra. Note that the GMEB allows
to include a cosmological constant term along $\tilde{\alpha}_{3}$.

At the level of the gauge fields one can see that the relativistic gauge
fields can be expressed in terms of the NR ones as follows%
\begin{eqnarray}
W^{0} &=&\omega +\frac{s}{2\xi ^{2}}\,,\text{\ \ \ \ \ }W^{a}=\frac{\omega
^{a}}{\xi }\,,\text{ \ \ \ \ \ \ \thinspace \thinspace \thinspace }S=\omega -%
\frac{s}{2\xi ^{2}}\,,  \notag \\
E^{0} &=&\xi \tau +\frac{m}{2\xi }\,,\text{ \ \ \ \ \thinspace \thinspace }%
E^{a}=e^{a}\,,\text{ \ \ \ \ \ \ \ \ }M=\xi \tau -\frac{m}{2\xi }\,,
\label{NRfields} \\
K^{0} &=&\xi ^{2}k+\frac{t}{2}\,,\text{ \ \ \ \ \ }K^{a}=\xi k^{a}\,,\text{
\ \ \ \ \ \ \ }T=\xi ^{2}k-\frac{t}{2}\,,  \notag \\
N^{0} &=&\xi ^{3}f+\xi \frac{v}{2},\text{ \ \ \thinspace \thinspace\ }%
N^{a}=\xi ^{2}f^{a},\text{ \ \ \ \ \ \ }V=\xi ^{3}f-\xi \frac{v}{2}  \notag
\end{eqnarray}%
in order to have that $A=\tilde{A}$. Then considering the rescaling of the
relativistic parameters as in (\ref{alphas}) and considering (\ref{NRfields}%
) in the relativistic CS action (\ref{U1CS}), we find the NR CS action (\ref%
{NRaction}) after applying the limit $\xi \rightarrow \infty $.

As an ending remark, one could consider, as in the Maxwell case, the
inclusion of three gauge fields in the relativistic generalized Maxwell
algebra and then study its NR version. Although a NR limit of the
[generalized Maxwell]$\oplus u\left( 1\right) ^{3}$ gravity theory could be
defined, it is possible to show that the respective NR gravity theory has a
degenerate bilinear form. Such feature would imply that the equations of
motion from such NR theory do not determine all the dynamical fields. Then,
in order to have well-defined field equations we need non-degenerate
invariant tensor which requires to consider a CS action based on the
[generalized Maxwell]$\oplus u\left( 1\right) ^{4}$ algebra as the
relativistic gravity theory. In particular, we have that the field equations
of the GMEB theory are given by the vanishing of each curvature (\ref{Curva}%
).

\section{Generalized family of non-relativistic algebras and semigroup
expansion}

The expansion of a Lie algebra is a method that consists in finding a new (bigger) Lie algebra $\mathfrak{G}$, following a series of well-defined steps from an original Lie algebra $\mathfrak{g}$. This procedure was first introduced by \cite{Hatsuda:2001pp} and later studied in \cite{AIPV,AIPV2}. It basically consists in performing a rescaling by a real parameter $\lambda$ of some of the coordinates of the Lie group $g^{i}, i=1,...,$ dim $\mathfrak{g}$, and then expanding the Maurer-Cartan (MC) one-forms in powers of the parameter $\lambda$. An alternative expansion method, called as $S$-expansion, was subsequently introduced in \cite{Sexp}. The $S$-expansion procedure consists in obtaining a new Lie algebra $%
\mathfrak{G}=S\times \mathfrak{g}$, by combining the elements of a
semigroup $S$ with the structure constants of a Lie algebra $\mathfrak{g}$. This approach is entirely based on operations performed on the algebra generators, whereas the aforesaid power series expansion is carried out on the MC one-forms. Another point in which both procedures differ, lies in the fact that the $S$-expansion is defined on the Lie algebra $\mathfrak{g}$ without mentioning the group manifold, while the power series expansion is based on the rescaling of the group coordinates. Remarkably, the semigroup expansion method can reproduce the MC forms power series expansion for a particular choice of the semigroup $S$. On the other hand, one of the advantage of working with the $S$-expansion is that it not only provides the commutation relations of the expanded algebra, but also allows to compute the non-vanishing components of the invariant tensor of the expanded algebra in terms of the invariant tensor for the original algebra.

In this section, following the procedure used in \cite{CR4, PSR}, we review a generalized family of NR algebras by considering the semigroup expansion method. Then, we extend the results obtained in
\cite{PSR} by showing that the $S$-expansion not only allows to obtain
expanded NR algebras but also provides with their relativistic versions and
the appropriate rescaling of the generators allowing a proper NR limit.

Here, we consider the Nappi-Witten algebra \cite{NW, FFSS} as the original
algebra $\mathfrak{g}$, which can be seen as a central extension of the
homogeneous part of the Galilei algebra. The Nappi-Witten algebra is spanned
by the set of generators $\left\{ \tilde{J},\tilde{G}_{a},\tilde{S}\right\} $
which satisfy the following non-vanishing commutation relations,%
\begin{eqnarray}
\left[ \tilde{J},\tilde{G}_{a}\right]  &=&\epsilon _{ab}\tilde{G}_{b}\,,
\notag \\
\left[ \tilde{G}_{a},\tilde{G}_{b}\right]  &=&-\epsilon _{ab}\tilde{S}\,,
\label{NW}
\end{eqnarray}%
where $\tilde{J}$ are spatial rotations, $\tilde{G}_{a}$ are Galilean boosts
and $\tilde{S}$ is a central charge. Such algebra can be obtained as a
contraction of an $U\left( 1\right) $-enlargement of the Lorentz algebra
through the identification of the [Lorentz]$\oplus u\left( 1\right) $
generators with the Nappi-Witten ones as%
\begin{equation}
J_{0}=\frac{\tilde{J}}{2}+\xi ^{2}\tilde{S}\,,\text{\ \ \ \ \ }J_{a}=\xi
\tilde{G}_{a}\,,\text{ \ \ \ \ \ \ }Y=\frac{\tilde{J}}{2}-\xi ^{2}\tilde{S}%
\,.  \label{limit}
\end{equation}%
Then the Nappi-Witten algebra is obtained after applying the limit $\xi
\rightarrow \infty $. One can see that the non-vanishing components of a
non-degenerate invariant tensor of the Nappi-Witten algebra read%
\begin{eqnarray}
\left\langle \tilde{J}\tilde{S}\right\rangle  &=&-1\,,  \notag \\
\left\langle \tilde{G}_{a}\tilde{G}_{b}\right\rangle  &=&\delta _{ab}\,.
\label{NWinvt}
\end{eqnarray}

In what follows, we shall review the family of NR symmetries obtained as $%
S_{E}^{\left( N\right) }$-expansions of the Nappi-Witten algebra \cite{PSR}.
We shall denote such family as generalized extended Bargmann GEB$^{\left(
N\right) }$ algebra. Interestingly, we will see that the extended Bargmann,
the MEB and the GMEB algebra previously introduced in the previous section
are particular cases of the GEB$^{\left( N\right) }$\ algebra. Furthermore,
we will show that the same semigroup can be used at the relativistic level
providing with the relativistic version of the GEB$^{\left( N\right) }$
algebra. Before approaching the family of NR algebras and their respective
NR\ CS\ actions, we first show that the GMEB algebra can alternatively be
obtained by expanding the Nappi-Witten algebra.

\subsection{Generalized Maxwellian exotic Bargmann gravity from semigroup
expansion}

Let $S_{E}^{\left( 3\right) }=\left\{ \lambda _{0},\lambda _{1},\lambda
_{2},\lambda _{3},\lambda _{4}\right\} $ be the relevant semigroup whose
elements satisfy the following multiplication law%
\begin{equation}
\lambda _{\alpha }\lambda _{\beta }=\left\{
\begin{array}{lcl}
\lambda _{\alpha +\beta }\,\,\,\, & \mathrm{if}\,\,\,\,\alpha +\beta <4\,, &
\\
\lambda _{4}\,\, & \mathrm{if}\,\,\,\,\alpha +\beta \geq 4\,, &
\end{array}%
\right.  \label{ml}
\end{equation}%
being $\lambda _{4}=0_{S}$ the zero element, such that $0_{S}\lambda
_{\alpha }=0_{S}$. An expanded algebra can be obtained after performing a $%
0_{S}$-reduction to the $S_{E}^{\left( 3\right) }$-expansion of the
Nappi-Witten algebra following the definitions of \cite{Sexp}. The expanded
generators $\left\{ \tilde{J},\tilde{G}_{a},\tilde{H},\tilde{P}_{a},\tilde{Z}%
,\tilde{Z}_{a},\tilde{N},\tilde{N}_{a},\tilde{S},\tilde{M},\tilde{T},\tilde{V%
}\right\} $ are related to the Nappi-Witten ones through the semigroup
elements as%
\begin{equation}
\begin{tabular}{c|ccc}
$\lambda _{3}$ & $\tilde{N}$ & $\tilde{N}_{a}$ & $\tilde{V}$ \\
$\lambda _{2}$ & $\tilde{Z}$ & $\tilde{Z}_{a}$ & $\tilde{T}$ \\
$\lambda _{1}$ & $\tilde{H}$ & $\tilde{P}_{a}$ & $\tilde{M}$ \\
$\lambda _{0}$ & $\tilde{J}$ & $\tilde{G}_{a}$ & $\tilde{S}$ \\ \hline
& $\tilde{J}$ & $\tilde{G}_{a}$ & $\tilde{S}$%
\end{tabular}%
\end{equation}%
Let us note that the $0_{S}$-reduction condition implies that $0_{S}T_{A}=0$%
, with $T_{A}$ being a generator of the original algebra. Using the
commutation relations of the Nappi-Witten algebra (\ref{NW}) and the
multiplication law of the semigroup (\ref{ml}), one can show that the
expanded generators satisfy the GMEB\ algebra (\ref{NRgen}) previously
introduced.

Interestingly, the $S$-expansion procedure can also provide with the
non-vanishing components of the invariant tensor for the expanded algebra.
Indeed, considering the Theorem VII of \cite{Sexp}, it is possible to
express the invariant tensor of the expanded algebra in terms of the
Nappi-Witten ones (\ref{NWinvt}) as%
\begin{equation}
\left\langle T_{A}^{\left( \nu \right) }T_{B}^{\left( \mu \right)
}\right\rangle _{S_{E}^{\left( 3\right) }\times \mathfrak{g}}=\alpha
_{\gamma }K_{\nu \mu }^{\quad \gamma }\left\langle T_{A}T_{B}\right\rangle _{%
\mathfrak{g}}\,,
\end{equation}%
where $T_{A}^{\left( \nu \right) }$ are the corresponding expanded
generators, $\alpha _{\gamma }$ are arbitrary constants and $K_{\nu \mu
}^{\quad \gamma }$ is the 2-selector for $S_{E}^{\left( 3\right) }$
satisfying%
\begin{equation}
K_{\nu \mu }^{\quad \gamma }=\left\{
\begin{array}{ll}
1,\,\,\,\, & \mathrm{when}\,\,\,\,\gamma =\nu +\mu \text{ \textrm{and} }\nu
+\mu <4\text{ } \\
0,\, & \mathrm{otherwise.}%
\end{array}%
\right.
\end{equation}%
Then one can see that the non-vanishing components of the invariant tensor
for the expanded algebra are those of the GMEB algebra given by (\ref{nrinv}%
). Thus, the $S_{E}^{\left( 3\right) }$-expansion of the Nappi-Witten
algebra provides not only with the commutation relations of the GMEB algebra
but also with its invariant tensor which is the crucial ingredient for the
construction of a CS action.

\subsection{Generalized Extended Bargmann family}

A family of generalized extended Bargmann algebra can be obtained by $S$%
-expanding the Nappi-Witten algebra (\ref{NW}) considering $S_{E}^{\left(
N\right) }=\left\{ \lambda _{0},\lambda _{1},\lambda _{2},\dots ,\lambda
_{N,}\lambda _{N+1}\right\} $ as the relevant semigroup \cite{PSR}. In
particular, the elements of the semigroup $S_{E}^{\left( N\right) }$ satisfy
the following multiplication law%
\begin{equation}
\lambda _{\alpha }\lambda _{\beta }=\left\{
\begin{array}{lcl}
\lambda _{\alpha +\beta }\,\,\,\, & \mathrm{if}\,\,\,\,\alpha +\beta <N+1 &
\\
\lambda _{N+1}\,\, & \mathrm{if}\,\,\,\,\alpha +\beta \geq N+1 &
\end{array}%
\right.   \label{ml2}
\end{equation}%
with $\lambda _{N+1}\equiv 0_{S}$ being the zero element of the semigroup
such that $0_{S}\lambda _{\alpha }=0_{S}$. An expanded algebra is found by
performing a $0_{S}$-reduction to the $S_{E}^{\left( N\right) }$-expansion
of the Nappi-Witten algebra. The expanded NR\ algebra is spanned by the set
of the generators $\left\{ \tilde{J}^{\left( i\right) },\tilde{G}%
_{a}^{\left( i\right) },\tilde{S}^{\left( i\right) }\right\} $ with $%
i=0,\dots ,N$. Such NR generators are related to the Nappi-Witten ones
through the semigroup elements as%
\begin{equation}
\tilde{J}^{\left( i\right) }=\lambda _{i}\tilde{J},\qquad \tilde{G}%
_{a}^{\left( i\right) }=\lambda _{i}\tilde{G}_{a}\,,\qquad \tilde{S}^{\left(
i\right) }=\lambda _{i}\tilde{S}\,.
\end{equation}%
Then, using the multiplication law of the semigroup (\ref{ml2}) and the
original commutators of the Nappi-Witten algebra (\ref{NW}), one can show
that the expanded NR algebra satisfy the following non-vanishing commutation
relations%
\begin{eqnarray}
\left[ \tilde{J}^{\left( i\right) },\tilde{G}_{a}^{\left( j\right) }\right]
&=&\epsilon _{ab}\tilde{G}_{b}^{\left( i+j\right) }\,,  \notag \\
\left[ \tilde{G}_{a}^{\left( i\right) },\tilde{G}_{b}^{\left( j\right) }%
\right]  &=&-\epsilon _{ab}\tilde{S}^{\left( i+j\right) }\,,  \label{GEBN}
\end{eqnarray}%
for $i+j<N+1\,$.The expanded NR\ algebra can be seen as a generalization of
the extended Bargmann algebra and is denoted as GEB$^{\left( N\right) }$
algebra.

Interestingly, one can show that the extended Bargmann algebra \cite{BR} is
obtained for $N=1$. Indeed, we have that the GEB$^{\left( 1\right) }$
algebra, which is given by%
\begin{eqnarray}
\left[ \tilde{J}^{\left( 0\right) },\tilde{G}_{a}^{\left( 0\right) }\right]
&=&\epsilon _{ab}\tilde{G}_{b}^{\left( 0\right) }\,,\qquad \left[ \tilde{G}%
_{a}^{\left( 0\right) },\tilde{G}_{b}^{\left( 0\right) }\right] =-\epsilon
_{ab}\tilde{S}^{\left( 0\right) }\,,  \notag \\
\left[ \tilde{J}^{\left( 0\right) },\tilde{G}_{a}^{\left( 1\right) }\right]
&=&\epsilon _{ab}\tilde{G}_{b}^{\left( 1\right) }\,,\qquad \left[ \tilde{G}%
_{a}^{\left( 0\right) },\tilde{G}_{b}^{\left( 1\right) }\right] =-\epsilon
_{ab}\tilde{S}^{\left( 1\right) }\,,  \label{EB} \\
\left[ \tilde{J}^{\left( 1\right) },\tilde{G}_{a}^{\left( 0\right) }\right]
&=&\epsilon _{ab}\tilde{G}_{b}^{\left( 1\right) }\,,  \notag
\end{eqnarray}%
corresponds to the extended Bargmann algebra by identifying the generators as%
\begin{eqnarray}
\tilde{J}^{\left( 0\right) } &=&\tilde{J}\,,\quad \tilde{G}_{a}^{\left(
0\right) }=\tilde{G}_{a}\,,\quad \tilde{S}^{\left( 0\right) }=\tilde{S}\,,
\notag \\
\tilde{J}^{\left( 1\right) } &=&\tilde{H}\,,\quad \tilde{G}_{a}^{\left(
1\right) }=\tilde{P}_{a}\,,\quad \tilde{S}^{\left( 1\right) }=\tilde{M}\,.
\label{id1}
\end{eqnarray}

On the other hand, for $N=2$, one can see that the GEB$^{\left( 2\right) }$
algebra corresponds to the MEB algebra introduced in \cite{AFGHZ}, which is
given by (\ref{EB}) along with%
\begin{eqnarray}
\left[ \tilde{J}^{\left( 0\right) },\tilde{G}_{a}^{\left( 2\right) }\right]
&=&\epsilon _{ab}\tilde{G}_{b}^{\left( 2\right) }\,,\qquad \left[ \tilde{G}%
_{a}^{\left( 0\right) },\tilde{G}_{b}^{\left( 2\right) }\right] =-\epsilon
_{ab}\tilde{S}^{\left( 2\right) }\,,  \notag \\
\left[ \tilde{J}^{\left( 2\right) },\tilde{G}_{a}^{\left( 0\right) }\right]
&=&\epsilon _{ab}\tilde{G}_{b}^{\left( 2\right) }\,,\qquad \left[ \tilde{G}%
_{a}^{\left( 1\right) },\tilde{G}_{b}^{\left( 1\right) }\right] =-\epsilon
_{ab}\tilde{S}^{\left( 2\right) }\,,  \label{MEB} \\
\left[ \tilde{J}^{\left( 1\right) },\tilde{G}_{a}^{\left( 1\right) }\right]
&=&\epsilon _{ab}\tilde{G}_{b}^{\left( 1\right) }\,,  \notag
\end{eqnarray}%
where one can consider the identification (\ref{id1}) together with%
\begin{equation}
\tilde{J}^{\left( 2\right) }=\tilde{Z}\,,\quad \tilde{G}_{a}^{\left(
2\right) }=\tilde{Z}_{a}\,,\quad \tilde{S}^{\left( 2\right) }=\tilde{T}\,.
\label{id2}
\end{equation}

The GMEB algebra presented here can also be seen as a particular case of the
GEB$^{\left( N\right) }$ algebra. In fact, as we have previously shown, the
GMEB algebra appears as a $S_{E}^{\left( 3\right) }$-expansion of the
Nappi-Witten algebra.

Interestingly, the GEB$^{\left( N\right) }$ algebra is the respective NR
version of $U\left( 1\right) $-enlargements of the so-called $\mathfrak{B}%
_{N+2}$ algebra introduced in \cite{EHTZ, GRCS},%
\begin{equation*}
\mathfrak{B}_{N+2}\oplus u\left( 1\right) ^{N+1}=\left\{
\begin{array}{l}
\text{\lbrack Poincaré]}\oplus u\left( 1\right) ^{2}\,\  \\
\text{\lbrack Maxwell]}\oplus u\left( 1\right) ^{3}\, \\
\text{\lbrack Gen. Maxwell]}\oplus u\left( 1\right) ^{4} \\
\vdots \\
\mathfrak{B}_{N+1}\oplus u\left( 1\right) ^{N} \\
\mathfrak{B}_{N+2}\oplus u\left( 1\right) ^{N+1}%
\end{array}%
\right. \quad \overset{\text{NR limit}}{\longrightarrow }\quad \text{GEB}%
^{\left( N\right) }=\left\{
\begin{array}{l}
\text{Extended Bargmann} \\
\text{MEB}\, \\
\text{GMEB} \\
\vdots \\
\text{GEB}^{\left( N-1\right) } \\
\text{GEB}^{\left( N\right) }%
\end{array}%
\right.
\end{equation*}%
Let us note that the Poincaré, Maxwell and generalized Maxwell algebras are
the $\mathfrak{B}_{3}$, $\mathfrak{B}_{4}$ and $\mathfrak{B}_{5}$ algebras,
respectively. Moreover, analogously to \cite{CR4}, the $S_{E}^{\left(
N\right) }$ semigroup used to obtain the GEB$^{\left( N\right) }$ algebra is
the same used to find the $\mathfrak{B}_{N+2}$ algebra from the Lorentz
algebra. Such particularity also appears for infinite-dimensional
(super)algebras \cite{CCRS, CCFR1, CCFR2}\ and algebras coupled to spin-3
\cite{CCFRS}. It is interesting to point out that the number of additional $%
U\left( 1\right) $ generators appearing in the relativistic algebra is
related to the $N+1$ elements of the semigroup $S_{E}^{\left( N\right) }$.
This is due to the fact that the $\mathfrak{B}_{N+2}\oplus u\left( 1\right)
^{N+1}$ algebra can be recovered as a $S_{E}^{\left( N\right) }$-expansion
of the [Lorentz]$\oplus u\left( 1\right) $ algebra.

Let us note that the $S$-expansion procedure can also provides with the
proper NR\ limit (\ref{con}) leading to the GEB$^{\left( N\right) }$ algebra
in terms of the contraction process (\ref{limit}) allowing to obtain the
Nappi-Witten algebra. As we have previously mentioned, the Nappi-Witten
algebra can be found as an IW contraction of the [Lorentz]$\oplus u\left(
1\right) $ algebra. Interestingly, the identification of the relativistic
generators defining $\mathfrak{B}_{N+2}\oplus u\left( 1\right) ^{N+1}$
algebra, with the NR\ generators (denoted with a tilde) can be defined as%
\begin{equation}
J_{0}^{\left( i\right) }=\frac{\tilde{J}^{\left( i\right) }}{2\xi ^{i}}+\xi
^{2-i}\tilde{S}^{\left( i\right) }\,,\text{\ \ \ \ \ }J_{a}^{\left( i\right)
}=\xi ^{1-i}\tilde{G}_{a}^{\left( i\right) }\,,\text{ \ \ \ \ \ \ }Y^{\left(
i\right) }=\frac{\tilde{J}^{\left( i\right) }}{2}-\xi ^{2-i}\tilde{S}%
^{\left( i\right) }\,,
\end{equation}%
with $i=0,1,\dots ,N$. Here, the relativistic generators are related to the
[Lorentz]$\oplus u\left( 1\right) $ ones through the elements of $%
S_{E}^{\left( N\right) }$ as%
\begin{equation}
J_{0}^{\left( i\right) }=\lambda _{i}J_{0}\,,\qquad J_{a}^{\left( i\right)
}=\lambda _{i}J_{a}\,,\qquad Y^{\left( i\right) }=\lambda _{i}Y\,.
\end{equation}%
Thus, the semigroup $S_{E}^{\left( N\right) }$ leads to the proper $U\left(
1\right) $-enlargement of the relativistic theory which leads to a
non-degenerate and finite NR gravity theory. One can check that for $N=1,2,3$
we recover the contraction process for the extended Bargmann, MEB and GMEB,
respectively.

On the other hand, as we have previously noted, an additional advantage of
the $S$-expansion method is that it provides with the invariant tensors of
the expanded algebra which are crucial for the construction of a CS action.
Thus, following Theorem VII of \cite{Sexp}, one can show that the
non-vanishing components of the invariant tensor of the GEB$^{\left(
N\right) }$ algebra is given by%
\begin{eqnarray}
\left\langle \tilde{J}^{\left( i\right) }\tilde{S}^{\left( i\right)
}\right\rangle  &=&-\tilde{\alpha}_{i+j}\,,  \notag \\
\left\langle \tilde{G}_{a}^{\left( i\right) }\tilde{G}_{b}^{\left( j\right)
}\right\rangle  &=&\tilde{\alpha}_{i+j}\,\delta _{ab}\,,  \label{ginvt}
\end{eqnarray}%
for $i+j<N+1$. \ Then the NR\ CS action based on the GEB$^{\left( N\right) }$
algebra expressed in term of the gauge connection one-form $A=\omega
^{\left( i\right) }\tilde{J}^{\left( i\right) }+\omega ^{a\left( i\right) }%
\tilde{G}_{a}^{\left( i\right) }+s^{\left( i\right) }S^{\left( i\right) }$
is given by%
\begin{equation}
I_{NR}=\int \tilde{\alpha}_{i}\,\left[ \omega _{a}^{\left( j\right) }d\omega
^{a\left( k\right) }\delta _{j+k}^{i}+\epsilon ^{ac}\omega _{a}^{\left(
j\right) }\omega ^{\left( k\right) }\omega _{c}^{\left( l\right) }\delta
_{j+k+l}^{i}-2s^{\left( j\right) }d\omega ^{\left( k\right) }\delta
_{j+k}^{i}\right] \,,
\end{equation}%
with $i=0,1,\dots ,N$. The NR CS gravity action contains $i$ independent
sectors each one invariant under the GEB$^{\left( N\right) }$ algebra. In
particular, the term proportional to $\alpha _{0}$ corresponds to the NR
exotic gravity action. The terms proportional to $\alpha _{1}$ and $\alpha
_{2}$ are the extended Bargmann gravity \cite{BR, HLO} and MEB gravity \cite%
{AFGHZ} actions, respectively. The NR\ CS action for the GMEB algebra
previously defined appears along $\alpha _{3}$. On the other hand, for $%
3<i\leq N$, the additional gauge fields related to $J^{\left( i\right) }$, $%
G_{a}^{\left( i\right) }$ and $S^{\left( i\right) }$ appear explicitly along
$\tilde{\alpha}_{i}$, corresponding to the respective NR CS action for the
GEB$^{\left( i\right) }$ algebra. Let us notice that a general expression
for the CS action based on an expanded Nappi-Witten algebra for a semigroup $S$ has been presented in \cite{PSR}.

\section{Conclusions}

In this paper we have presented a generalization of the Maxwellian extended
Bargmann gravity introduced in \cite{AFGHZ}. We have explicitly shown that this NR symmetry, that we
have denoted as GMEB, can be obtained as an IW contraction of the [generalized
Maxwell]$\oplus u\left( 1\right) ^{4}$ algebra. To this end, we first presented the CS gravity action invariant under the generalized Maxwell algebra. Then, we constructed an $U(1)$-enlargement which is required to have a well-defined NR limit. Interestingly, the GMEB
gravity theory contains the MEB and the extended Bargmann theories as
sub-cases.  The GMEB algebra belongs to a generalized family of NR algebras which can be obtained by expanding the Nappi-Witten algebra \cite{PSR}. Here, we have shown that the expansion procedure based on
semigroups is a powerful tool in the NR context since it provides not
only with the commutation relations and invariant tensor of the expanded NR
algebras, but also with the respective relativistic algebra required to
obtain non-degenerate finite NR gravity theories.

Our results could be useful in the presence of supersymmetry. The
construction of proper NR supergravity models are non-trivial and have only
recently been approached. In particular the respective NR superalgebras have
mainly been constructed by hand in three spacetime dimensions \cite{ABRS,
BRZ, BR, OOTZ, CRR2, OOZ}. The expansion method considered here could not
only be used as an alternative and straightforward way to obtain known and
new NR superalgebras but also to construct NR supergravity actions.
Furthermore, it could bring invaluable information about the relativistic
versions and the respective NR limits as in our case. Let us notice that the
Lie algebra expansion method using the Maurer-Cartan equations \cite{AIPV,
AIPV2} has also been used in the NR\ context with diverse interesting
results \cite{BIOR, AGI, Romano, KOOZ}.

Let us note that, as was shown in \cite{PSR}, the GEB$^{(N)}$ algebra appears as a IW contraction of another family of NR algebras denoted as generalized Newton-Hooke. Then, another aspect that deserved to be explored is the derivation of this generalized Newton-Hooke algebra as an IW contraction of a family of relativistic algebras. One could conjecture that, similarly to our results, the semigroup procedure could be useful to elucidate the appropriate $U(1)$-enlargement of the relativistic family.

Regarding the relativistic generalized Maxwell gravity theory, it would be
interesting to explore its general solution and asymptotic symmetry. In
particular, one could analyze the influence of the additional gauge field in
the vacuum energy and angular momentum, and compare them to those of General
Relativity \cite{BGG, MOR} and usual Maxwell theory \cite{CMMRSV}. One could
expect that the asymptotic structure is given by the infinite-dimensional
enhancement of the $\mathfrak{B}_{5}$ algebra introduced in \cite{CCRS}.

\section*{Acknowledgements}

This work was supported by the National Agency for Research and Development ANID (ex-CONICYT) - PAI grant No. 77190078 (P.C.), FONDECYT Projects N$^{\circ }$3180594 (M.I.) and N$^{\circ }$3170438
(E.R.). P.C. would like to thank to the Dirección de Investigación and
Vice-rectoría de Investigación of the Universidad Católica de la Santísima
Concepción, Chile, for their constant support.


\begin{thebibliography}{99}
\bibitem{Kuchar} K. Kuchar, \textit{Gravitation, geometry, and
nonrelativistic quantum theory}, Phys. Rev. D \textbf{22} (1980) 1285.%
\textit{\ }

\bibitem{DK} C. Duval, H.P. Kunzle, \textit{Minimal Gravitational Coupling
in the Newtonian Theory and the Covariant Schrödinger Equation}, Gen. Rel.
Grav. \textbf{16} (1984) 333.

\bibitem{DBKP} C. Duval, G. Burdet, H.P. Kunzle, M. Perrin, \textit{Bargmann
Structures and Newton-Cartan Theory}, Phys. Rev. D \textbf{31} (1985) 1841.

\bibitem{DGH} C. Duval, G.W. Gibbons, P. Horvathy, \textit{Celestial
mechanics, conformal structures and gravitational waves}, Phys. Rev. D%
\textbf{43} (1991) 3907. [hep-th/0512188].

\bibitem{Duval} C. Duval, \textit{On Galilean isometries}, Class. Quant.
Grav. \textbf{10} (1993) 2217. arXiv:0903.1641 [math-ph].

\bibitem{DLP} R. De Pietri, L. Lusanna, M. Pauri, \textit{Standard and
generalized Newtonian gravities as
\'{}%
gauge%
\'{}
theories of the extended Galilei group. I. The standard theory}, Class.
Quant. Grav. \textbf{12} (1995) 219. [gr-qc/9405046].

\bibitem{DLP2} R. De Pietri, L. Lusanna, M. Pauri, \textit{Standard and
generalized Newtonian gravities as
\'{}%
gauge%
\'{}
theories of the extended Galilei group. II. Dynamical three space theories},
Class. Quant. Grav. \textbf{12} (1995) 255. [gr-qc/9405047].

\bibitem{Horava} P. Ho\v{r}ava, \textit{Quantum Gravity at a Lifshitz Point}%
, Phys. Rev. D \textbf{79} (2009) 084008. arXiv:0901.3775 [hep-th].

\bibitem{DH} C. Duval, P.A. Horvathy,\textit{\ Non-relativistic conformal
symmetries and Newton-Cartan structures}, J. Phys. A \textbf{42} (2009)
465206. arXiv:0904.0531 [math-ph].

\bibitem{ABPR} R. Andringa, E. Bergshoeff, S. Panda, M. de Roo, \textit{%
Newtonian Gravity and the Bargmann Algebra}, Class. Quant. Grav. \textbf{28}
(2011) 105011. arXiv:1011.1145 [hep-th].

\bibitem{ABGR} R. Andringa, E. Bergshoeff, J. Gomis, M. de Roo, \textit{%
\'{}%
Stringly%
\'{}
Newton-Cartan Gravity}, Class. Quant. Grav. \textbf{29} (2012) 235020.
arXiv:1206.5176 [hep-th].

\bibitem{ABRS} R. Andringa, E.A. Bergshoeff, J. Rosseel, E. Sezgin, \textit{%
3D Newton-Cartan supergravity}, Class. Quantum Gravity \textbf{30 }(2013)
205005. arXiv:1305.6737 [hep-th].

\bibitem{BMM} R. Banerjee, A. Mitra, P. Mukherjee, \textit{Localisation of
the Galilean symmetry and dynamical realisation of Newton-Cartan geometry},
Class. Quant. Grav. \textbf{32} (2015) 045010. arXiv:1407.3617 [hep-th].

\bibitem{BM1} X. Bekaert, K. Morand, \textit{Connections and dynamical
trajectories in generalised Newton-Cartan gravity I. An intrinsic view}, J.
Math. Phys. \textbf{57} (2016) 022507. arXiv:1412.8213 [hep-th].

\bibitem{BM2} X. Bekaert, K. Morand, \textit{Connections and dynamical
trajectories in generalised Newton-Cartan gravity II. An ambient perspective}%
, J. Math. Phys. \textbf{59} (2018) 072503. arXiv:1505.03739 [hep-th].

\bibitem{BRZ2} E. Bergshoeff, J. Rosseel, T. Zojer, \textit{Newton-Cartan
supergravity with torsion and Schrödinger supergravity}, JHEP \textbf{1511}
(2015) 180. arXiv:1509.04527 [hep-th].

\bibitem{BMu} R. Banerjee, P. Mukherjee, \textit{Torsional Newton-Cartan
geometry from Galilean gauge theory}, Class. Quant. Grav. \textbf{33} (2016)
225013. arXiv:1604.06893 [gr-qc].

\bibitem{BCRR} E. Bergshoeff, A. Chatzistavrakidis, L. Romano, J. Rosseel,
\textit{Newton-Cartan Gravity and Torsion}, JHEP \textbf{1710} (2017) 194.
arXiv:1708.05414 [hep.th].

\bibitem{HHO} D. Hansen, J. Hartong, N.A. Obers, \textit{Action principle
for Newtonian gravity}, Phys. Rev. Lett. \textbf{122} (2019) 061106.
arXiv:1807.04765 [hep-th].

\bibitem{BMu2} R. Banerjee, P. Mukherjee, \textit{Galilean gauge theory from
Poincare gauge theory}, Phys. Rev. D \textbf{98} (2018) 124021.
arXiv:1810.03902 [gr-qc].

\bibitem{OOTZ} N. Ozdemir, M. Ozkan, O. Tunca, U. Zorba, \textit{%
Three-Dimensional Extended Newtonian (Super)Gravity}, JHEP \textbf{1905}
(2019) 130. arXiv:1903.09377 [hep-th].

\bibitem{MPS} J. Matulich, S. Prohazka, J. Salzer, \textit{Limits of
three-dimensional gravity and metric kinematical Lie algebras in any
dimension}, JHEP \textbf{07} (2019) 118. arXiv:1903.09165 [hep-th].

\bibitem{CS} D. Chernyavsky, D. Sorokin, \textit{Three-dimensional
(higher-spin) gravities with extended Schrödinger and l-conformal Galilean
symmetries}, JHEP \textbf{07} (2019) 156. arXiv:1905.13154 [hep-th].

\bibitem{CRR3} P. Concha, L. Ravera, E. Rodríguez, \textit{Three-dimensional
exotic Newtonian gravity with cosmological constant}, Phys. Lett. B \textbf{%
804} (2020) 135392. arXiv:1912.02836 [hep-th].

\bibitem{CRR2} P. Concha, L. Ravera, E. Rodríguez, \textit{Three-dimensional
Maxwellian extended Bargmann supergravity}, JHEP \textbf{04} (2020) 051. arXiv:1912.09477 [hep-th].

\bibitem{GKPSR} J. Gomis, A. Kleinschmidt, J. Palmkvist, P.
Salgado-Rebolledo,\textit{\ Newton-Hooke/Carrollian expansions of (A)dS and
Chern-Simons gravity}, JHEP \textbf{2002} (2020) 009. arXiv:1912.07564
[hep-th].

\bibitem{BGSR} E. Bergshoeff, J. Gomis, P. Salgado-Rebolledo, \textit{%
Non-relativistic limits and three-dimensional coadjoint Poincare gravity},
arXiv:2001.11790 [hep-th].

\bibitem{EHVdB} M. Ergen, E. Hamamci, D. Van den Bleeken, \textit{Oddity in
nonrelativistic, strong gravity}, arXiv:2002.02688 [gr-qc].

\bibitem{AT} A. Achucarro, P.K. Townsend, \textit{A Chern-Simons action for
Three-dimensional anti-De Sitter Supergravity Theories}, Phys. Lett. B
\textbf{180} (1986) 89.

\bibitem{Witten} E. Witten, \textit{(2+1)-Dimensional gravity as an exactly
soluble system}, Nucl. Phys. B \textbf{311} (1988) 46.

\bibitem{Zanelli} J. Zanelli, \textit{Lecture notes on Chern-Simons
(super-)gravities}. Second edition (February 2008), [hep-th/0502193].

\bibitem{BRZ} E. Bergshoeff, J. Rosseel, T. Zojer, \textit{Newton-Cartan
(super)gravity as a non-relativistic limit}, Class. Quant. Grav. \textbf{32}
(2015) 205003. arXiv:1505.02095 [hep-th].

\bibitem{GO} J. Gomis, H. Ooguri, \textit{Nonrelativistic closed string
theory}, J. Math. Phys. \textbf{42} (2001) 3127. [hep-th/0009181].

\bibitem{BCG} A. Barducci, R. Casalbuoni, J. Gomis, \textit{Non-relativistic
Spinning Particle in a Newton-Cartan Background}, JHEP\textbf{\ 01} (2018)
002. arXiv:1710.10970 [hep-th].

\bibitem{BR} E.A. Bergshoeff, J. Rosseel, \textit{Three-Dimensional Extended
Bargmann Supergravity}, Phys. Rev. Lett. \textbf{116 }(2016) 251601.
arXiv:1604.08042 [hep-th].

\bibitem{HLO} J. Hartong, Y. Lei, N.A. Obers, \textit{Nonrelativistic
Chern-Simons theories and three-dimensional Horava-Lifshitz gravity}, Phys.
Rev. D \textbf{94} (2016) 065027. arXiv:1604.08054 [hep-th].

\bibitem{PS} G. Papageorgiou, B.J. Schroers, \textit{Galilean quantum
gravity with cosmological constant and the extended q-Heisenberg algebra},
JHEP \textbf{11} (2010) 020. arXiv:1008.0279 [hep-th].

\bibitem{AMO} O. Arratia, M.A. Martin, M.A. Olmo, \textit{Classical systems
and representations of }$\left( 2+1\right) $\textit{\ Newton-Hooke symmetries%
}, [math-ph/9903013].

\bibitem{Gao} Y.H. Gao, \textit{Symmetries, matrices, and de Sitter gravity}%
, [hep-th/0107067].

\bibitem{GP} G.W. Gibbons, C.E. Patricot, \textit{Newton-Hooke spacetimes,
Hpp-waves and the cosmological constant}, Class. Quant. Grav. \textbf{20}
(2003) 5225. [hep-th/0308200].

\bibitem{BGK} J. Brugues, J. Gomis, K. Kamimura, \textit{Newton-Hooke
Algebras, Non-relativistic Branes and Generalized pp-wave Metrics}, Phys.
Rev. D \textbf{73} (2006) 085011. [hep-th/0603023].

\bibitem{AGKP} P.D. Alvarez, J. Gomis, K. Kamimura, M.S. Plyushchay, $\left(
2+1\right) $\textit{D exotic Newton-Hooke symmetry, duality and projective
phase}, Annals Phys. \textbf{322} (2007) 1556. [hep-th/0702014].

\bibitem{OOZ} N. Ozdemir, M. Ozkan, U. Zorba, \textit{Three-Dimensional
Extended Lifshitz, Schrödinger and Newton-Hooke Supergravity}, JHEP \textbf{%
1911} (2019) 052. arXiv:1909.10745 [hep-th].

\bibitem{AFGHZ} L. Avilés, E. Frodden, J. Gomis, D. Hidalgo, J. Zanelli,
\textit{Non-Relativistic Maxwell Chern-Simons Gravity}, JHEP \textbf{1805 }%
(2018) 047. arXiv:1802.08453 [hep-th].

\bibitem{Schrader} R. Schrader, \textit{The Maxwell group and the quantum
theory of particles in classical homogeneous electromagnetic fields},
Fortsch. Phys. \textbf{20} (1972) 701.

\bibitem{BCR} H. Bacry, P. Combe, J.L. Richard, \textit{Group-theoretical
analysis of elementary particles in an external electromagnetic field. 1.
The relativistic particle in a constant and uniform field}, Nuovo Cim. A
\textbf{67} (1970) 267.

\bibitem{GK} J. Gomis, A. Kleinschmidt, \textit{On free Lie algebras and
particles in electro-magnetic fields}, JHEP \textbf{07} (2017) 085.
arXiv:1705.05854 [hep-th].

\bibitem{SSV} P. Salgado, R.J. Szabo, O. Valdivia, \textit{Topological
gravity and transgression holography}, Phys. Rev. D\textbf{89} (2014)
084077. arXiv:1401.3653 [hep-th].

\bibitem{HR} S. Hoseinzadeh, A. Rezaei-Aghdam, \textit{(2+1)-dimensional
gravity from Maxwell and semisimple extension of the Poincaré gauge
symmetric models}, Phys. Rev. D\textbf{90} (2014) 084008. arXiv:1402.0320
[hep-th].

\bibitem{CFRS} P.K. Concha, O. Fierro, E.K. Rodríguez, P. Salgado, \textit{%
Chern-Simons supergravity in $D=3$ and Maxwell superalgebra}, Phys. Lett. B
\textbf{750} (2015) 117. arXiv:1507.02335 [hep-th].

\bibitem{CMMRSV} P. Concha, N. Merino, O. Miskovic, E. Rodríguez, P.
Salgado-Rebolledo, O. Valdivia, \textit{Asymptotic symmetries of
three-dimensional Chern-Simons gravity for the Maxwell algebra}, JHEP
\textbf{1810} (2018) 079. arXiv:1805.08834 [hep-th].

\bibitem{BS} S. Bansal, D. Sorokin, \textit{Can Chern-Simons or
Rarita-Schwinger ber a Volkov-Akulov Goldstone?}, JHEP \textbf{07 }(2018)
106. arXiv:1806.05945 [hep-th].

\bibitem{CSDS} D. Chernyavsky, N. Sadik Deger, D. Sorokin, \textit{%
Spontaneously broken 3d Hietarinta-Maxwell Chern-Simons Theory and Minimal
Massive Gravity}, arXiv:2002.07592 [hep-th].

\bibitem{EHTZ} J.D. Edelstein, M. Hassaine, R. Troncoso, J. Zanelli, \textit{%
Lie-algebra expansions, Chern-Simons theories and the Einstein-Hilbert
Lagrangian}, Phys. Lett. B \textbf{640} (2006) 278. [hep-th/0605174].

\bibitem{GRCS} F. Izaurieta, E. Rodríguez, P. Minning, P. Salgado, A. Perez,
\textit{Standard General Relativity from Chern-Simons Gravity}, Phys. Lett.
B \textbf{678} (2009) 213. arXiv:0905.2187 [hep-th].

\bibitem{CPRS1} P.K. Concha, D.M. Peñafiel, E.K. Rodríguez, P. Salgado,
\textit{Even-dimensional General Relativity from Born-Infeld gravity}, Phys.
Lett. B \textbf{725} (2013) 419. arXiv:1309.0062 [hep-th].

\bibitem{CPRS2} P.K. Concha, D.M. Peñafiel, E.K. Rodríguez, P. Salgado,
\textit{Chern-Simons and Born-Infeld gravity theories and Maxwell algebras
type}, Eur. Phys. J. C \textbf{74} (2014) 2741. arXiv:1402.0023 [hep-th].

\bibitem{CPRS3} P.K. Concha, D.M. Peñafiel, E.K. Rodríguez, P. Salgado,
\textit{Generalized Poincaré algebras and Lovelock-Cartan gravity theory},
Phys. Lett. B \textbf{742} (2015) 310. arXiv:1405.7078 [hep.th].

\bibitem{CCFRS} R. Caroca, P. Concha, O. Fierro, E. Rodríguez, P.
Salgado-Rebolledo,\textit{\ Generalized Chern-Simons higher-spin gravity
theories in three dimensions}, Nucl. Phys. B \textbf{934} (2018) 240.
arXiv:1712.09975 [hep-th].

\bibitem{IW} E. Inönü, E.P. Wigner, \textit{On the contraction of groups and
their representations}, Proc. Natl. Acad. Sci USA \textbf{39} (1953) 510.

\bibitem{WW} E. Weimar-Woods, \textit{Contractions, generalized Inönü-Wigner
contractions and deformations of finite-dimensional Lie algebras}, Rev. Mod.
Phys. \textbf{12} (2000) 1505.

\bibitem{Sexp} F. Izaurieta, E. Rodríguez, P. Salgado, \textit{Expanding Lie
(super)algebras through Abelian semigroups}, J. Math. Phys. \textbf{47}
(2006) 123512. [hep-th/0606215].

\bibitem{CKMN} R. Caroca, I. Kondrashuk, N. Merino, F. Nadal, \textit{%
Bianchi spaces and their 3-dimensional isometries as S-expansions of
2-dimensional isometries}, J. Phys. A. \textbf{46} (2013) 225201.
arXiv:1104.3541 [math-ph].

\bibitem{AMNT} L. Andrianopoli, N. Merino, F. Nadal, M. Trigiante, \textit{%
General properties of the expansion methods of Lie algebras}, J. Phys. A
\textbf{46} (2013) 365204. arXiv:1308.4832 [gr-qc].

\bibitem{ACIPS} M. Artebani, R. Caroca, M.C. Ipinza, D.M. Peñafiel, P.
Salgado, \textit{Geometrical aspects of the Lie algebra S-expansion procedure%
}, J. Math. Phys. \textbf{57} (2016) 023516. arXiv:1602.04525 [math-ph].

\bibitem{ILPR} M.C. Ipinza, F. Lingua, D.M. Peñafiel, L. Ravera, \textit{An
analytic method for S-expansion involving resonance and reduction}, Fortsch.
Phys. \textbf{64} (2016) 854. arXiv:1609.05042 [hep-th].

\bibitem{PSR} D.M. Peñafiel, P. Salgado-Rebolledo, \textit{Non-relativistic
symmetries in three space-time dimensions and the Nappi-Witten algebra},
Phys. Lett. B \textbf{798} (2019) 135005. arXiv:1906.02161 [hep-th].

\bibitem{SS} D.V. Soroka, V.A. Soroka,\textit{\ Semi-simple extension of the
(super) Poincaré algebra, }Adv. High Energy Phys. \textbf{2009} (2009)
234147. [hep-th/0605251].

\bibitem{DFIMRSV} J. Díaz, O. Fierro, F. Izaurieta, N. Merino, E. Rodriguez,
P. Salgado, O. Valdivia, \textit{A generalized action for }$\mathit{(2+1)}$%
\textit{-dimensional Chern-Simons gravity}, J. Phys. A. Math. Theor.\textbf{%
\ 45} (2012) 255207, arXiv:1311.2215 [gr-qc].

\bibitem{SaSa} P. Salgado, S. Salgado,$\mathfrak{so}(D-1,1)\oplus \mathfrak{%
so}(D-1,2)$\textit{\ algebras and gravity}, Phys. Lett. B\textbf{\ 728 }%
(2014) 5-10.

\bibitem{CPR} P. Concha, D.M. Peñafiel, E. Rodríguez, \textit{On the Maxwell
supergravity and flat limit in 2+1 dimensions}, Phys. Lett. B \textbf{785}
(2018) 247. arXiv:1807.00194 [hep-th].

\bibitem{Concha} P. Concha, $\mathcal{N}$\textit{-extended Maxwell
supergravities as Chern-Simons theories in three spacetime dimensions},
Phys. Lett. B \textbf{792} (2019) 290. arXiv:1903.03081 [hep-th].

\bibitem{CR4} P. Concha, E. Rodríguez, \textit{Non-relativistic gravity
theory based on an enlargement of the extended Bargmann algebra}, JHEP
\textbf{07} (2019) 085. arXiv:1906.00086 [hep-th].

\bibitem{Hatsuda:2001pp} M.~Hatsuda, M.~Sakaguchi, \textit{Wess-Zumino Term for the AdS Superstring and Generalized Inönü-Wigner Contraction}, Prog. Theor. Phys. \textbf{109} (2003) 853-867. [hep-th/0106114].

\bibitem{AIPV} J.A. de Azcárraga, J.M. Izquierdo, M. Picón, O. Varela,
\textit{Generating Lie and gauge free differential (super)algebras by
expanding Maurer-Cartan forms and Chern-Simons supergravity}, \newline
Nucl. Phys. B \textbf{662} (2003) 185. [hep-th/0212347].

\bibitem{AIPV2} J.A. de Azcárraga, J.M. Izquierdo, M. Picón, O. Varela,
\textit{Expansions of algebras and superalgebras and some applications},
Int. J. Theor. Phys. \textbf{46 }(2007) 2738. [hep-th/0703017].

\bibitem{NW} C.R. Nappi, E. Witten, \textit{A WZW model based on a
nonsemisimple group}, Phys. Rev. Lett. \textbf{71} (1993) 3751.
[hep-th/9310112].

\bibitem{FFSS} J. Figueroa-O'Farrill, S. Stanciu, \textit{More D-branes in the Nappi-Witten background},
JHEP \textbf{01} (2000) 024. [hep-th/9909164].

\bibitem{CCRS} R. Caroca, P. Concha, E. Rodríguez, P. Salgado-Rebolledo,
\textit{Generalizing the }$\mathfrak{bms}_{3}$\textit{\ and 2D-conformal
algebras by expanding the Virasoro algebra}, Eur. Phys. J. C \textbf{78}
(2018) 262. arXiv:1707.07209 [hep-th].

\bibitem{CCFR1} R. Caroca, P. Concha, O. Fierro, E. Rodríguez, \textit{%
Three-dimensional Poincaré supergravity and }$\mathcal{N}$\textit{-extended
supersymmetric BMS}$_{3}$ \textit{algebra}, Phys. Lett. B \textbf{792}
(2019) 93. arXiv:1812.05065 [hep-th].

\bibitem{CCFR2} R. Caroca, P. Concha, O. Fierro, E. Rodríguez, \textit{On
the supersymmetric extension of asymptotic symmetries in three spacetime
dimensions}, Eur. Phys. J. C \textbf{80} (2020) 29. arXiv:1908.09150
[hep-th].

\bibitem{BIOR} E. Bergshoeff, J. Izquierdo, T. Ortín, L. Romano, \textit{Lie
Algebra Expansions and Actions for Non-Relativistic Gravity},\ JHEP \textbf{%
08} (2019) 048. arXiv:1904.08304 [hep-th].

\bibitem{AGI} J.A. de Azcárraga, D. Gútiez, J.M. Izquierdo, \textit{Extended
}$D=3$\textit{\ Bargmann supergravity from a Lie algebra expansion}, Nucl.
Phys. B \textbf{946} (2019) 114706. arXiv:1904.12786 [hep-th].

\bibitem{Romano} L. Romano, \textit{Non-Relativistic Four Dimensional
p-Brane Supersymmetric Theories and Lie Algebra Expansion}, arXiv:1906.08220
[hep-th].

\bibitem{KOOZ} O. Kasikci, N. Ozdemir, M. Ozkan, U. Zorba, \textit{%
Three-Dimensional Higher-Order Schrödinger Algebras and Lie Algebra
Expansions}, JHEP \textbf{04} (2020) 067. arXiv:2002.03558 [hep-th].

\bibitem{BGG} G. Barnich, A. Gomberoff, H.A. Gonzalez, \textit{The flat
limit of three dimensional asymptotically anti-de Sitter spacetimes}, Phys.
Rev. D\textbf{86} (2012) 024020. arXiv:1204.3288 [hep-th].

\bibitem{MOR} O. Miskovic, R. Olea, D. Roy, \textit{Vacuum energy in
asymptotically flat 2+1 gravity}, Phys. Lett. B \textbf{767} (2017) 258.
arXiv:1610.06101 [hep-th].
\end{thebibliography}
\end{document}